%% file: charm2015_Luchinsky.tex
\documentclass[12pt]{article}
\usepackage{graphicx}

\newcommand\pubnumber{arXiv:1503.00246 [hep-ph]}
\newcommand\pubdate{\today}

\def\ihep{
SRC IHEP of NRC "Kurchatov Institute"
}
\def\itep{
NRC "Kurchatov Institute" FSBI "SSC RF ITEP"
}
\def\support{\footnote{The work was carried out with the financial support of FRRC and RFBR (grant \# 4-02-00096 A)}}

\def\Title#1{\begin{center} {\Large #1 } \end{center}}
\def\Author#1{\begin{center}{ \sc #1} \end{center}}
\def\Address#1{\begin{center}{ \it #1} \end{center}}

\newcommand\pubblock{\rightline{\begin{tabular}{l} \pubnumber\\
         \pubdate  \end{tabular}}}
\newenvironment{Abstract}{\begin{quotation}  }{\end{quotation}}
\newenvironment{Presented}{\begin{quotation} \begin{center} 
             PRESENTED AT\end{center}\bigskip 
      \begin{center}\begin{large}}{\end{large}\end{center} \end{quotation}}

\input econfmacros.tex

\begin{document}
\begin{titlepage}
\pubblock

\vfill
\Title{$J/\Psi \Upsilon$ Production at LHC}
\vfill
\Author{A.V. Luchinsky\support}
\Address{\ihep\\ \itep}
\vfill
\begin{Abstract}
Inclusive production of $\Upsilon J/\psi$ pair in proton-proton interation at LHCb is considered. This process is forbidden at leading order of perturbation theory, so such channels as double parton scattering, $\chi_b\chi_c$ pair production with subsequent radiative decays of $P$-wave quarkonia, contributions of color-octet states, and NLO corrections are studied in details. For all these channels we present theoretical predictions of total cross sections at LHCb and distributions over different kinematical variables. According to presented in the paper results, double parton interaction gives main contribution to the cross section of the considered reaction.\end{Abstract}
\vfill
\begin{Presented}
The 7th International Workshop on Charm Physics (CHARM 2015)\\
Detroit, MI, 18-22 May, 2015
\end{Presented}
\vfill
\end{titlepage}
\def\thefootnote{\fnsymbol{footnote}}
\setcounter{footnote}{0}

\newcommand{\Q}{\mathcal{Q}}
\newcommand{\OS}{O_S}
\newcommand{\OP}{O_P}
\newcommand{\Br}{\mathrm{Br}}
\newcommand{\MM}{\mathcal{M}}
\newcommand{\A}{\mathcal{A}}
\newcommand{\B}{\mathcal{B}}
\newcommand{\GeV}{\mathrm{GeV}}
\newcommand{\SPS}{\mathrm{SPS}}
\newcommand{\fb}{\mathrm{fb}}
\newcommand{\pb}{\mathrm{pb}}
\newcommand{\NLO}{\mathrm{NLO}}
\newcommand{\CO}{\mathrm{CO}}
\newcommand{\DPS}{\mathrm{DPS}}
\newcommand{\mub}{\mathrm{\mu b}}
\newcommand{\mb}{\mathrm{mb}}
\newcommand{\nb}{\mathrm{nb}}

\section{Introduction}

Inclusive double production of heavy quarkonia was observed in NA3 experiment, where production of two $J/\psi$ mesons in $\pi N$ interaction was studied. The cross section of double $J/\psi$ production was suppressed by about three orders of magnitude in comparison with single $J/\psi$. Such suppression can be easily explained at leading order (LO) of perturbative QCD \cite{Kartvelishvili:1984ur,Humpert:1983yj}. In low energy hadronic collisions quark-antiquark subprocess  $q\overline{q}\to J/\psi J/\psi$ gives main contribution to the considered reaction, while in the case of high energy interactions gluonic subprocesses $gg\to J/\psi J/\psi$ are dominant. Leading order $O(\alpha_s^4)$ QCD predictions and subsequent next to leading order (NLO) corrections \cite{Lansberg:2013qka} are in  good agreement with experimental results, obtained at LHC \cite{Aaij:2014rms, Berezhnoy:2012xq, Berezhnoy:2011xy}. It is clear, on the other hand, that at LHC energies the 
gluoinc luminosity is large, so production of two $J/\psi$ mesons in two independent partonic interactions (so called double parton scattering, DPS) is possible. In papers \cite{Baranov:2011ch, Novoselov:2011ff} it is shown, that cross sections calculated in DPS approximation are comparable with the ones obtained in single parton scattering (SPS) approximation. The drawback of DPS is poor knowledge of double parton distribution functions. It turns out that while for double $J/\psi$ production DPS and SPS cross sections are comparable, in the case of $\Upsilon(1S)\Upsilon(1S)$ final state DPS cross section is suppressed by more than order of magnitude.

Question concerning situation in $\Upsilon J/\psi$ production arises naturally. At leading $O(\alpha_s^4)$ order in color-singlet approximation this process is forbidden, so the following relation holds:
\begin{eqnarray}
\sigma_{\SPS}^{\Upsilon J/\psi} & \ll & \sigma_{\SPS}^{\Upsilon\Upsilon}\ll\sigma_{\SPS}^{J/\psi J/\psi}.\label{eq:sigmaRel}
\end{eqnarray}
The reason is that at leading $O(\alpha_s^4)$ order for $J/\psi J/\psi$ and $\Upsilon\Upsilon$ processes diagrams with $t$-channel quark exchange are allowed, that saturate low invariant mass region and lead to rapid decrease of the cross section with the increase of this invariant mass, $\hat\sigma\sim1/\hat{s}^{2}$. At the same order of magnitude the contributions to $\Upsilon J/\psi$ production are absent, but production via radiative decays of $\chi_b\chi_c$ state is allowed. 

The rest of the paper will be organised as follows. In section \ref{sec:partonic} the analytical expressions for cross sections of partonic reactions $gg\to\Q_1\Q_2$ are presented. Cross sections of hadronic reactions $pp\to \chi_b \chi_c +X$, $pp\to J/\psi \Upsilon+X$ are given in section \ref{sec:hadronic}. The effect of double parton scattering is also discussed in this section. The results of our paper are briefly summarised in the last section.

\section{Partonic Cross Sections\label{sec:partonic}}

\begin{figure}[htb]
\centering
\includegraphics[width=0.9\textwidth]{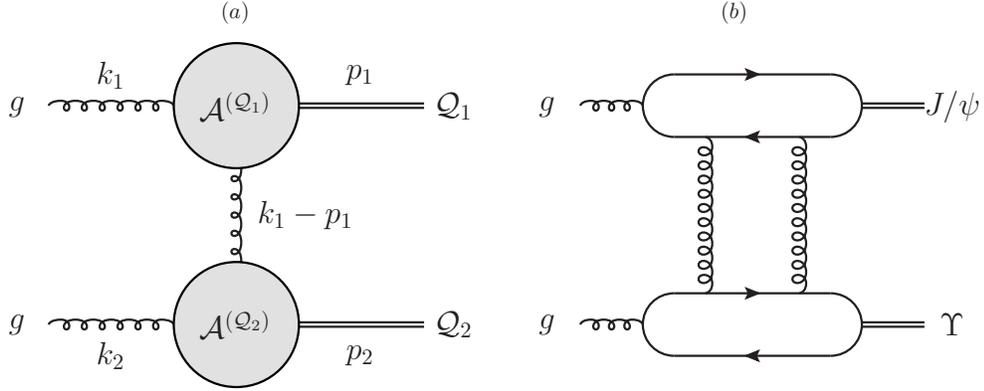}
\caption{Typical Feynman diagrams for $gg\to\chi_b\chi_c$ process (left figure) $gg\to J/\psi\Upsilon$ at NLO (right diagram)}
\label{fig:diag}
\end{figure}

It is well known that production of $J/\psi \Upsilon$ pair in $gg$ integration in colour-singlet approximation at leading order is forbidden. Final vector quarkonia particles, however, can be produced via radiative decays of $P$-wave quarkonia $\chi_{b,c}$, that can be produced at LO:
\begin{eqnarray}
d\hat\sigma(gg\to J/\psi \Upsilon + X) = \sum_{J_b,J_c=0}^2 \Br\left(\chi_{cJ_c}\to J/\psi\gamma\right) \Br\left(\chi_{bJ_b}\to\Upsilon\gamma\right) \hat\sigma(gg\to\chi_{bJ_b}\chi_{cJ_c}).
\end{eqnarray}
Typical Feynman diagram of this reaction is shown in Fig.\ref{fig:diag}(a), and the corresponding matrix element can be written in the form
\begin{eqnarray}
\MM(gg\to\chi_{bJ_b}\chi_{cJ_c}) &=& \frac{1}{\hat t}\A^{(\chi_c)}_{\alpha\mu}(k_1,p_1-k_1) \A^{(\chi_b)}_{\beta\mu}(k_2,p_1-k_1) \epsilon_1^\alpha \epsilon_2^\beta + \mathrm{permutations}.
\label{eq:amp}
\end{eqnarray}
In the above expression $\epsilon_{1,2}$ are polarisation vectors of initial gluons with moment $k_{1,2}$ and we have introduced the notation $\A^{\Q}_{\mu\nu}$(p,k) for the vertex of $\Q$ quarkonia production in fusion of two (probably virtual) gluons with momenta $p$, $k$ (see Fig.\ref{fig:diag}(b) for the example of the diagram).

Due to the presented in the amplitude (\ref{eq:amp}) propagator of virtual gluons main contributions to the cross sections of the considered reaction should come from small $\hat t$ kinematic region and in the case of high energy of the partonic reaction $\sqrt{\hat s}$ this gluon can become almost real. In this approximation one can obtain estimates for cross sections of scalar and tensor quarkonia production in high energy limit (see also \cite{Kiselev:1988mc}):
\begin{eqnarray*}
\hat{\sigma}\left(gg\to\chi_{bJ_{b}}\chi_{cJ_{c}}\right) & \sim & \left(2J_{c}+1\right)\left(2J_{b}+1\right)\frac{\Gamma\left(\chi_{cJ_{c}}\to2g\right)}{M_{\chi_{c}}}\frac{\Gamma\left(\chi_{bJ_{b}}\to2g\right)}{M_{\chi_{b}}}.
\end{eqnarray*}
The ratios of these cross sections are equal to
\begin{eqnarray}
\hat{\sigma}\left(\chi_{c2}\chi_{b2}\right):\hat{\sigma}\left(\chi_{c0}\chi_{b2}\right):\hat{\sigma}\left(\chi_{c2}\chi_{b0}\right):\hat{\sigma}\left(\chi_{c0}\chi_{b0}\right) & = & \frac{16}{9}:\frac{4}{3}:\frac{4}{3}:1.\label{eq:ratios}
\end{eqnarray}
If there is at least one axial meson in the final state, the corresponding cross sections should be suppressed.

\begin{figure}
\includegraphics[width=\textwidth]{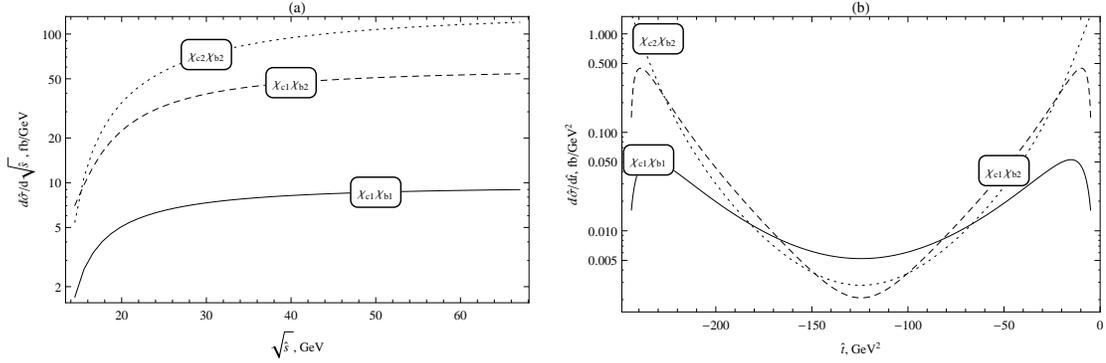}
\caption{
Dependence of partonic cross sections on $\sqrt{\hat s}$ (left panel) and $\hat t$ dependence at $'hat s=\hat s_0$  (right panel) for  $gg\to\chi_{b}\chi_{c}$ reactions. On both figures solid, dashed and dotted lines correspond to $gg\to\chi_{c1}\chi_{b1}$, $gg\to\chi_{c1}\chi_{b2}$, and  $gg\to\chi_{c2}\chi_{b2}$ subprocesses respectively.
\label{fig:hSigma}}
\end{figure}

Presented in papers \cite{Likhoded:2013aya,Likhoded:2014gpa,Likhoded:2014kfa} analysis, however, shows, that experimental data are better described with a larger value
\begin{eqnarray*}
\left|R_{cc}'(0)\right|^{2} & = & 0.35\pm0.05\ \GeV^{5}.
\end{eqnarray*}
For bottomonium mesons the same approach gives \cite{Likhoded:2012hw}
\begin{eqnarray*}
\left|R'_{bb}(0)\right|^{2} & \approx & 1.0\ \GeV^{5}.
\end{eqnarray*}
These numbers will be used in the following. We will also neglect the fine splitting of heavy quarkonia masses, the following values will be used:
\begin{eqnarray*}
M_{\chi_{c}} & = & 3.5\ \GeV,\qquad M_{\chi_{b}}=9.9\ \GeV.
\end{eqnarray*}
The scale of the strong coupling constant $\alpha_{s}(\mu^{2})$ is chosen to be
\begin{eqnarray*}
\mu^{2} & = & \frac{M_{\chi_{c}}^{2}+M_{\chi_{b}}^{2}}{2}.
\end{eqnarray*}
Using presented above values it is easy to obtain predictions for partonic cross sections of $\chi$ meson production in gluon-gluon interaction. Integrated over total phase space partonic cross sections of the considered reactions at energy
\begin{eqnarray}
\hat{s} & = & \hat{s}_{0}=2(M_{\chi_{c}}+M_{\chi_{b}})^{2},\label{eq:s0}
\end{eqnarray}
for example, are listed in table \ref{tab:hSigma}. It is clearly seen that for ratios of scalar and tensor mesons production cross sections relations (\ref{eq:ratios}) are satisfied with pretty good accuracy. Dependence of some of these cross sections on $\sqrt{\hat s}$ and their distributions over $\hat{t}$ are shown in Fig.\ref{fig:hSigma}. One can see, that in the case of tensor meson production in low $\hat{t}$, $\hat{u}$ regions an increase caused by $\hat{t}$-channel gluon is observed. For axial meson production, on the other hand, in agreement with Landau-Yang theorem such increase is absent.

\begin{figure}
\includegraphics[width=1\textwidth]{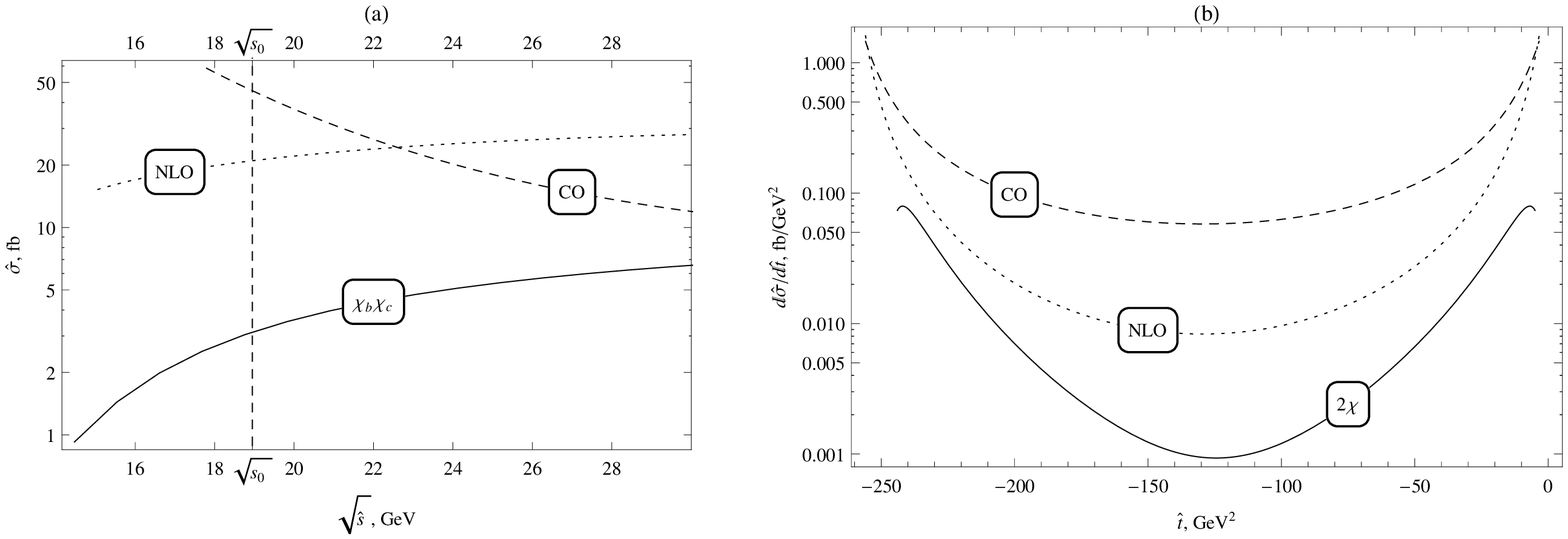}
\caption{
Dependence of cross section of $\Upsilon J/\psi$ production in partonic subprocess (left panel) and its $\hat t$-distribution at $\hat s=\hat{s}_0$ (right panel). Solid, dotted, and dashed lines correspond to $\chi_b\chi_c$, NLO, and color-octet channels respectively.\label{fig:hplotNLO}}
\end{figure}

\begin{table}
\begin{center}
\begin{tabular}{|c|c|c|c|}
\hline 
$\Q_{1}/\Q_{2}$ & $\chi_{c0}$ & $\chi_{c1}$ & $\chi_{c2}$\tabularnewline
\hline 
\hline 
$\chi_{b0}$ & 16.3 & 14.8 & 21.4\tabularnewline
\hline 
$\chi_{b1}$ & 2.1 & 4.6 & 3.8\tabularnewline
\hline 
$\chi_{b2}$ & 21.4 & 19.6 & 29.2\tabularnewline
\hline 
\end{tabular}
\caption{Total cross sections of the partonic reactions $gg\to\chi_{cJ_{c}}\chi_{bJ_{b}}$ (in fb) at $\hat{s}=\hat{s}_{0}$\label{tab:hSigma}}
\end{center}
\end{table}

In the Introduction it was mentioned already that prohibition of color singlet $J/\psi\Upsilon$ production can be bypassed if one consider NLO corrections (see typical diagram shown in Fig.\ref{fig:diag}). In the case of $J/\psi$ pair production this effect was considered in paper \cite{Kiselev:1988mc} (see also \cite{Ginzburg:1988zy}), and it is simple to extend it to $J/\psi\Upsilon$ case. The cross section of the considered reaction can be written in the form
\begin{eqnarray}
\frac{d\hat{\sigma}(gg\to\Upsilon J/\psi)}{d\hat{t}} & = & \frac{25\pi^{3}\alpha_{s}^{6}f_{\psi}^{2}f_{\Upsilon}^{2}}{46656m_{b}^{4}m_{c}^{4}}\left[\frac{F(\hat{t})+F(\hat{u})}{2}\right]^{2},\label{eq:dsdtNLO}
\end{eqnarray}
where
\begin{eqnarray}
F(t) & = & \frac{m_{b}m_{c}}{\pi}\int\frac{d^{2}k}{\mathbf{k}^{2}(\mathbf{k}-\mathbf{q})^{2}}\left\{ \frac{1}{1+\tau_{c}}-\frac{1}{1+\tau_{c}\mathbf{r}^{2}}\right\} \left\{ \frac{1}{1+\tau_{b}}-\frac{1}{1+\tau_{b}\mathbf{r}^{2}}\right\} \label{eq:F}
\end{eqnarray}
and notations
\begin{eqnarray*}
\tau_{b,c} & = & \frac{\mathbf{q}^{2}}{4m_{b,c}^{2}},\qquad\mathbf{r}=\frac{2\mathbf{k}-\mathbf{q}}{|\mathbf{q}|}.
\end{eqnarray*}
are introduced. One can easily see that after substitution $m_b=m_c$, $e_b=e_c$ these results agree with those presented in ref.\cite{Ginzburg:1988zy}. 
The mesonic constants in (\ref{eq:dsdtNLO}) are equal to $f_{\psi}  =  0.4\ \GeV$, $f_{\Upsilon}=0.66\ \GeV$. Using these parameters we have obtained the number
\begin{eqnarray*}
\hat{\sigma}_{\SPS}^{NLO}\left(gg\to\Upsilon J/\psi\right) & = & 21\ \fb.
\end{eqnarray*}
for integrated partonic cross section at $\hat{s}=\hat{s}_{0}$.
Dependence of this cross section on $\sqrt{\hat s}$ and its $\hat t$ distribution at $\hat s=\hat{s}_0$ are shown in Fig.\ref{fig:hplotNLO} by dotted line. It should be noted however, that these results were obtained in the limit $\hat{s}\gg|\hat{t}|\gg M_{\psi,\Upsilon}^{2},$, so they can be treated only as rough estimates. In our future works we are going to remove this limitation and obtain predictions for shown in Fig.\ref{fig:diag}(b) cross sections in full kinematical region.

The other way to bypass prohibition of $J/\psi \Upsilon$ production in $gg$ interaction is to consider color octet components of final vector quarkonia \cite{Ko:2010xy}. Using presented in this paper analytical results and
\begin{eqnarray*}
\left\langle O_{1}^{J/\psi}\right\rangle  & = & 1.3\ \GeV^{3},\qquad\qquad\left\langle O_{1}^{\Upsilon}\right\rangle =9.2\ \GeV^{3},\\
\left\langle O_{8}^{J/\psi}\right\rangle  & = & 3.9\times10^{-3}\ \GeV^{3},\qquad\left\langle O_{8}^{\Upsilon}\right\rangle =0.15\ \GeV^{3},
\end{eqnarray*}
for NRQCD matrix elements \cite{Bodwin:2007fz,Kang:2007uv,Braaten:1999qk,Kramer:2001hh}, it is easy to obtain distribution shown by dashed line in Fig.\ref{fig:hplotNLO}. According to this figure $\sqrt{\hat s}$ dependence of different cross sections is quite different, with CO channel giving dominant contribution in region of small invariant mass of the final pair.

\section{Hadronic Cross Sections\label{sec:hadronic}}

Let us now consider cross section of hadronic production of $J/\Psi \Upsilon$ pair in hadronic collisions at LHC. Using presented in the previous section results for partonic cross sections it is easy to obtain the following values for hadronic ones:
\begin{eqnarray*}
\sigma_{\SPS}\left(pp\to\chi_{b}\chi_{c}+X\to\Upsilon J/\psi+X\right) & = & 0.2\ \pb,\\
\sigma_{\SPS}\left(pp\to\Upsilon J/\psi+X,\NLO\right) & = & 1.5\ \pb,\\
\sigma_{\SPS}\left(pp\to\Upsilon J/\psi,\CO\right) & = & 11.1\ \pb.
\end{eqnarray*}

It is well known, that at LHC energies double parton scattering (DPS) could also give noticeable contribution to cross section of the considered reaction. In this model final vector quarkonia are produced in two independent partonic subprocesses and the cross section of this reaction can be written in surprisingly simple form:
\begin{eqnarray*}
\sigma_{\DPS}\left(\Upsilon J/\psi\right) & = & \frac{\sigma_{\SPS}(\Upsilon)\sigma_{\SPS}(J/\psi)}{\sigma_{eff}}.
\end{eqnarray*}
Using experimental results  $\sigma_{\SPS}(\Upsilon)=0.14\ \mub$, $\sigma_{\SPS}(J/\psi)=1.28\ \mub$ for SPS cross sections and \cite{Aaij:2013yaa} and the value $\sigma_{eff}=14\ \mb$ \cite{Snigirev:2003cq,Korotkikh:2004bz,Novoselov:2011ff}, one can obtain the result
\begin{eqnarray*}
\sigma_{\DPS}\left(pp\to\Upsilon J/\psi+X\right) & = & 12.5\ \pb.
\end{eqnarray*}

\begin{figure}
\includegraphics[width=0.5\textwidth]{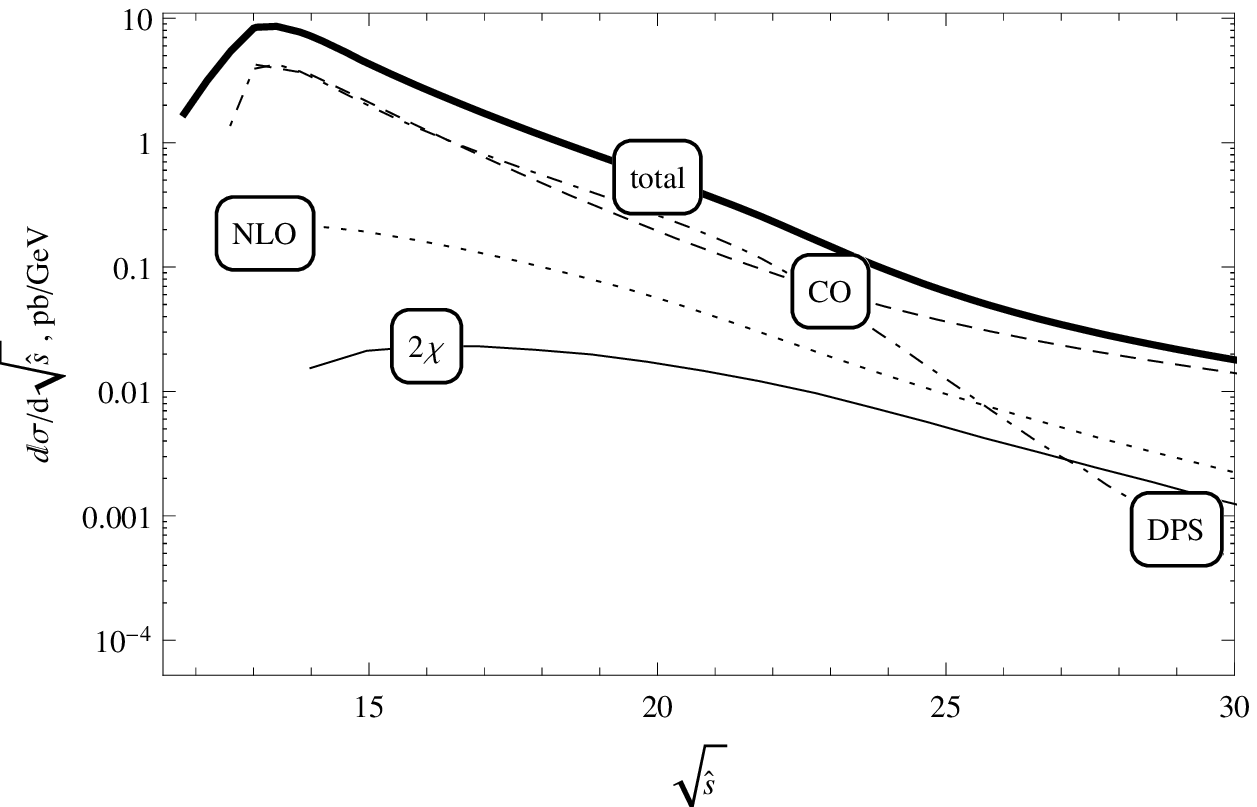}
\includegraphics[width=0.5\textwidth]{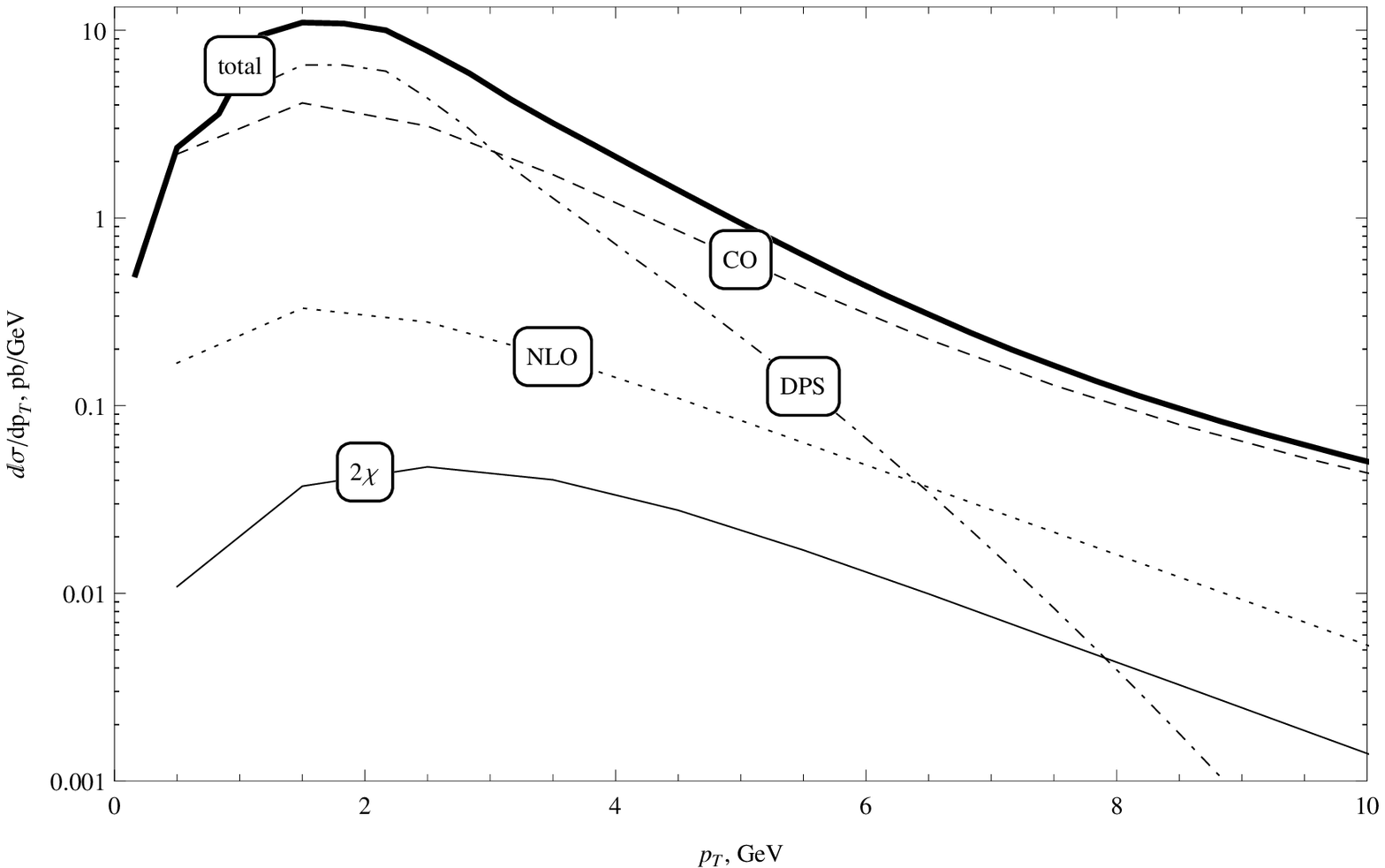}
\caption{
Distribution of $pp\to\Upsilon J/\psi+X$ reaction cross section at LHCb over $\Upsilon J/\psi$-pair invariant mass (left figure) and transverse momentum (right figure). Thin solid, dashed, dotted, and dash-dotted  lines correspond to contributions of $\chi_b\chi_c$, CO, NLO, and DPS channels respectively, while thick solid line stands for total cross section\label{fig:mass}}
\end{figure}

It is clear that this simple mechanism gives main contribution to the cross section, but one can separate contributions of other channels with the help of various kinematical distributions (see. Figs.\ref{fig:mass})

\section{Conclusion\label{sec:conclusion}}

Presented article is devoted to theoretical analysis of inclusive production of heavy quarkonia pair $\Upsilon J/\psi$ at LHCb. It is clear, that at leading order of perturbation theory in colour singlet approximation this process is forbidden, so other mechanisms should be studied. In our paper we consider such channels as double parton scattering, next to leading order corrections, colour octet model and production of vector quarkonia states via radiative decays of $\chi_{b,c}$ mesons. For all these processes theoretical predictions of total cross sections under LHCb detector conditions and distributions over kinematical variables are presented.

According to obtained in our article results (see also \cite{Likhoded:2015zna}), double parton scattering gives main contribution to the cross section of the considered process. From analysis of different distributions. however, it is also possible to separate contributions of other channel, so one can, for example, determine the value of colour-octet NRQCD matrix elements. So, we think that additional theoretical and experimental analysis of inclusive $\Upsilon J/\psi$ pair production at LHC is very interesting and actual task.

The author would like to thank A.K. Likhoded, S.V. Poslavsky and Dr. Belyaev for fruitful discussions. The work was carried out with the financial support of FRRC and RFBR (grant \# 4-02-00096 A)

\end{document}

%% file: econfmacros.tex
\def\beq{\begin{equation}}
\def\eeq#1{\label{#1}\end{equation}}
\def\eeqn{\end{equation}}

\def\beqa{\begin{eqnarray}}
\def\eeqa#1{\label{#1}\end{eqnarray}}
\def\eeqan{\end{eqnarray}}

\let\bar=\overbar

\def\Dslash{\not{\hbox{\kern-4pt $D$}}}
\def\dslash{\not{\hbox{\kern-2pt $\del$}}}

\def\msb{{\bar{\ssstyle M \kern -1pt S}}}

